\documentstyle[aaspp4]{article}
\singlespace

\def\ms{m\thinspace s$^{-1}$\ }

\def\deg{$^\circ$}
\def\etal{\it et al.}

\slugcomment{To appear in Astrophysical Journal Letters}

\lefthead{Hatzes et al.}
\righthead{A Planetary Companion to Epsilon Eridani}
\begin{document}
\title{Evidence for a Long-period Planet Orbiting Epsilon Eridani
\footnote{Based on observations collected at McDonald
Observatory, Lick Observatory, European Southern Observatory at La Silla,
and the Canada-France-Hawaii  Telescope}}
\author{Artie P. Hatzes$^1$,  William D. Cochran$^1$, Barbara McArthur$^1$,
Sallie L. Baliunas$^2$,
Gordon A.H. Walker$^3$, Bruce Campbell$^4$, Alan W. Irwin$^5$, 
Stephenson Yang$^5$,
Martin K\"urster$^6$, Michael Endl$^{6,7}$, Sebastian Els$^{6,8}$,
 R. Paul Butler$^9$ and Geoffrey W. Marcy$^{10}$}

\affil{$^1$McDonald Observatory, The University of Texas at Austin,
    Austin, TX 78712}

\affil{$^2$Harvard-Smithsonian Center for Astrophysics,
60 Garden Street, MS 15, Cambridge, MA 02138}

\affil{$^3$ Physics and Astronomy Department, University of
British Columbia, Vancouver, B.C. Canada V6T 1Z4}

\affil{$^4$ BTEC Enterprises Ltd. }

\affil{$^5$Department of Physics and Astronomy, University of Victoria, 
Victoria, BC, Canada, V8W 3P6 }

\affil{$^6$European Southern Observatory, Casilla 19001, Vitacura,
Santiago 19, Chile}

\affil{$^7$ Institut f\"ur Astronomie, Universit\"at Wien,
T\"urkenschanzstr.~17, A-1180, Austria}

\affil{$^8$ Institut  f\"ur Theoretische Astrophysik, Universit\"at
Heidelberg, Tiergartenstr.~15, D-69121 Heidelberg, Germany}

\affil{$^9$ Department of Terrestrial Magnetism, Carnegie Institution
of Washington, 5243 Broad Branch Road NW, Washington, DC 20015-1305. }

\affil{$^{10}$ Department of Astronomy, University of California,
Berkeley, CA 94720}

Subject headings: planetary systems -- stars: individual ($\epsilon$ Eridani)
-- stars: low-mass, brown dwarfs -- techniques: radial

\begin{abstract}
High precision radial velocity (RV) measurements spanning the years
1980.8--2000.0 are presented for the nearby (3.22 pc) K2 V star
$\epsilon$ Eri. These data, which  represent a combination of six independent
data sets taken with four different telescopes, show convincing variations
with a period of $\approx$ 7 yrs.   A least squares orbital solution using
robust estimation yields orbital parameters of period, $P$ = 6.9 yrs,
velocity $K$-amplitude $=$ 19 {\ms}, eccentricity $e$ $=$ 0.6, projected
companion  mass $M$ sin $i$ = 0.86 $M_{Jupiter}$, and semi-major axis
$a_2$ $=$ 3.3 AU. Ca II H\&K S-index measurements spanning the same time
interval show significant variations with periods of 3 and 20 yrs, yet none
at the RV period. If magnetic activity were responsible for the RV variations
then it produces a significantly different period than is seen in the Ca II
data. Given the lack of Ca II variation with the same period as that found
in the RV measurements, the long-lived and coherent nature of these
variations, and the high eccentricity of the implied orbit, Keplerian
motion due to a planetary companion seems to be the most likely explanation 
for the observed RV variations.  The wide angular  separation of the planet 
from the star  (approximately 1 arc-second) and the long orbital period
make this planet a prime candidate for both direct imaging and space-based
astrometric measurements.
\end{abstract}

\section{Introduction}

	Over the past 5 years radial velocity (RV) surveys have had  stunning
success at finding the first giant gaseous planets 
($M$ = 0.5 -- 10 $M_{Jupiter}$)
in orbit around other stars. Although to date over 40 of these systems
have been found, none
qualify as `true Jupiters', i.e. giant planets in low eccentricity
orbits with semimajor axes in excess of 3 AU. 

Epsilon Eri ($=$ HR 1084)  is a bright ($V$ = 3.7), nearby (3.22 pc)
K2 V star exhibiting a high level of chromospheric 
activity (e.g. Gray \& Baliunas 1995) that is consistent with 
its relatively young age of $<$ 1 Gyr (Soderblom \& D\"appen 1989).
This star also has a dusty ring 60 AU from the star (Greaves et al. 1998). 

Epsilon Eri has  
been the subject of several  RV planet searches. Walker et al. (1995) using
measurements spanning 11 years found evidence
for $\approx$ 10 yr variation with an amplitude of 15 {\ms}. These
results were substantiated by Nelson \& Angel (1998) using an analysis
of the same data set. Cumming et al. (1999)
analyzed 11 years of RV data on this star taken at Lick Observatory 
and found significant variations with comparable amplitude  
but with a shorter period of 6.9 years. Because of the high
level of magnetic activity for $\epsilon$~Eri
these RV variations were largely interpreted as arising
from a stellar activity cycle.

	The McDonald Observatory Planet Search Program 
(Cochran \& Hatzes 1999) has
been monitoring $\epsilon$~Eri  since late 1988.
In this paper these results in combination with other
surveys are presented.  We confirm the presence of long period RV variations
and demonstrate that they  likely owe to the presence of
a planetary companion.

\section{ The Radial Velocity Data Sets}

The RV data represent measurements from four   planet search groups.
Data taken from late 
1980 to late 1991 using the coud{\'e} spectrograph of the Canada-France-Hawaii
3.6m telescope (``CFHT'' set) utilized  an HF absorption 
cell for the velocity metric (Walker et al. 1995).
Cumming et al. (1999) presented
precise radial velocity measurements for $\epsilon$~Eri taken with
an iodine gas absorption cell spanning 1987.69 -- 1999.99 using the
Hamilton Spectrograph of the 3m Shane
Telescope at Lick Observatory (``Lick'' set). 
Observations of $\epsilon$~Eri
spanning the time interval 1992.84--1998.02  were made as part 
of an ESO planet search program using
the 1.4m Coude Auxiliary telescope $+$ CES spectrograph at La Silla. 
Details of this program
can  be found in K\"urster et al. (2000) and Endl et al. (2000).

	The largest data set is based on observations made at the 2.7m telescope
at McDonald Observatory and comprises three subsets. From 
1988.74--1994.81 observations
of $\epsilon$~Eri were made using telluric O$_2$ as the wavelength
reference (``McD-O2'' set). 
From late 1990 to late 1998 an iodine absorption cell was used
as the velocity metric (``McD-I2'' set). 
Both of these data sets utilized the ``6-ft'' camera of the coude spectrograph
which isolated a single order of the echelle and thus had limited
wavelength coverage (9 {\AA}), but high spectral resolution
(resolving power, $R$ = $\lambda$/$\delta \lambda$ = 240,000).
Starting in 1998 the McDonald program switched to 
using the 2-d coude spectrograph  (``McD-2dC'' set)
of the 2.7m telescope (Tull et al. 1995). This   instrument allowed
us to use the full spectral region having usable I$_2$ lines
($\sim$ 5000--6100 {\AA}) but at a lower resolving power
($R$ = 60,000).  Analysis 
of the radial velocity derived from the the iodine (I$_2$)
cell technique including modeling of the instrumental profile as outlined
in Valenti et al. (1995).

	Table 1 summarizes the various data sets, the RV technique
employed, the time interval of the observations, the number of observations,
and the rms scatter, $\sigma$, of each set (from the fitted orbit,
see $\S$4 for details). The number of observations represent nightly averages.

\begin{deluxetable}{cclcc}
\tablewidth{25pc}
\tablecaption{The Data Sets}
\tablehead{
\colhead{Data Set}  &  \colhead{Coverage} & \colhead{Technique} & \colhead{N}
& \colhead{$\sigma_{RV}$} \\
\colhead{}  &  \colhead{(Year)} & \colhead{} & \colhead{}
& \colhead{(m\,s$^{-1}$)}}
\startdata
McD-O2               & 1988.74--1994.81 & Telluric     & 29      & 22.8  \nl
McD-I2               & 1990.78--1998.07 & Iodine Cell  & 46      & 16.5 \nl
McD-2DC              & 1998.69--2000.03 & Iodine Cell  &  9      & 11.4  \nl
CFHT                 & 1980.81--1991.88 & HF cell      & 51      & 14.5 \nl
LICK                 & 1987.69--1998.99 & Iodine Cell  & 62      & 14.0  \nl
ESO                  & 1992.84--1998.02 & Iodine Cell  & 28      & 13.9  \nl
\enddata
\end{deluxetable}

\section{Period Search}

\begin{figure}[ht]
\plotone{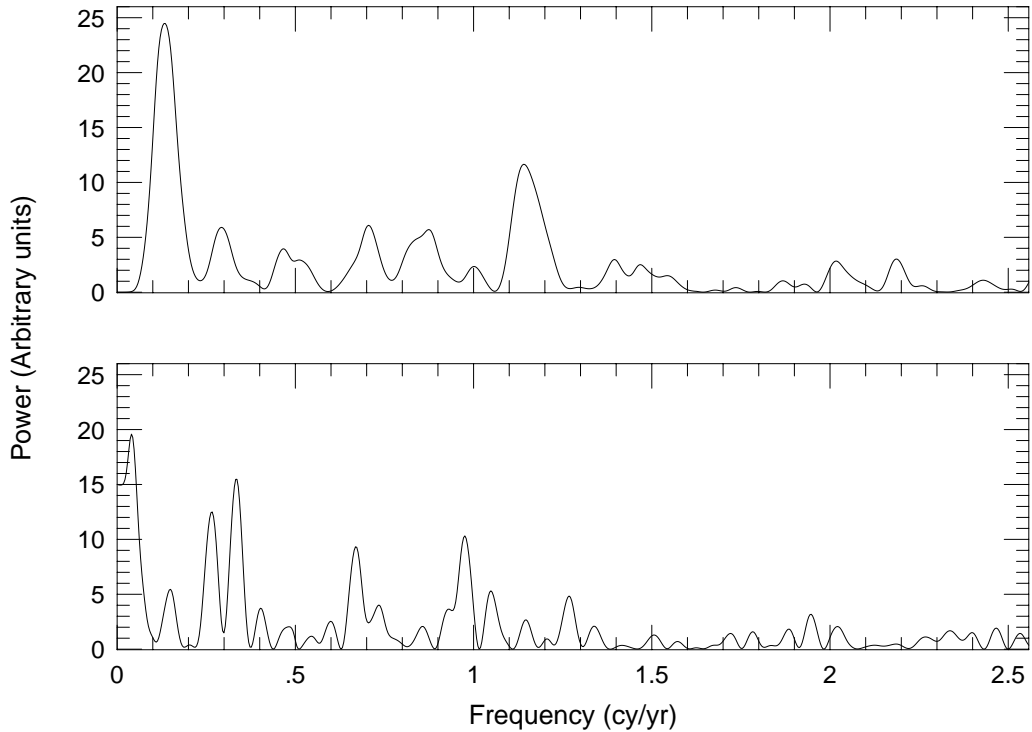}
\caption[ftall.ps]{(Top) The Lomb-Scargle periodogram of the combined
RV measurements for $\epsilon$~Eri.
(Bottom) Periodogram of the Ca II H\&K S-index measurements.
\label{periodogram}}
\end{figure}

	A Lomb-Scargle periodogram (Scargle 1982) 
was used to search for periodic signals in the data.   Each data set 
represents relative radial velocities that are 
have different zero-point offsets with respect to the others.
These offsets were determined in an unbiased manner 
from the orbital fitting  (see below) which
computed simultaneously all orbital elements  as well as the relative velocity
offsets (there is significant temporal overlap between the different sets).
The top panel
of Figure~\ref{periodogram} shows the resulting periodogram  of the combined
data which exhibits significant power at a period of 7.13 yr ($=$ 2603 d).
The false alarm probability (FAP) was estimated using Eq. 18 of Scargle (1982) 
and this yielded a FAP $=$ 5 $\times$ 10$^{-9}$.
This low  FAP was confirmed  using a bootstrap randomization scheme
(e.g. K\"urster et al. 1996).  
For comparison false data sets with the same variance as the original
observations were generated by randomly re-assigning the radial velocity
values the times of the observations.
This was done 5$\times 10^{5}$ times and there was no instance where maximum 
power in the ``random" periodogram exceeded that found in the data.

\section{The Orbital Solution}

An orbital solution was performed using a {\it GaussFit}
model (Jefferys et al. 1988; McArthur et al. 1994). The relative
velocity offsets of the different data sets were solved simultaneously
with the orbital elements and a long term trend
in the RV measurements. 
Both standard least squares and robust
estimation analyses were performed.   

\begin{figure}[ht]
\plotone{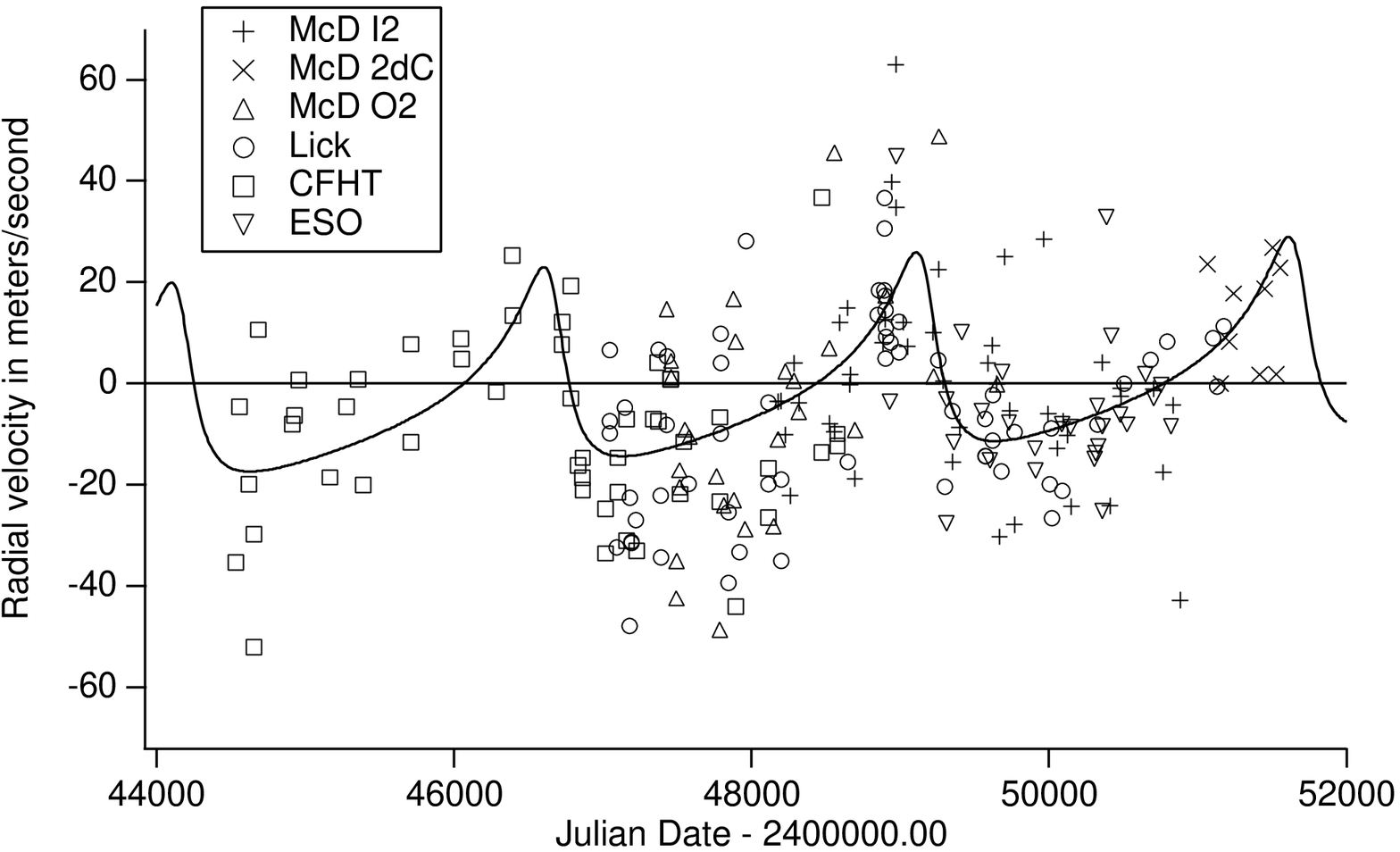}
\caption[orbit.ps]{Orbital solution to the 
RV data using the robust estimation elements of
Table 2. 
\label{orbit}}
\end{figure}

	Table 2 lists the orbital parameters  
from least squares
and robust estimation. The solutions in Table  2 include a slight slope
(0.4 m\,s$^{-1}$\,yr$^{-1}$) in the velocities. 
The orbital solution without the trend resulted in parameters 
consistent to  within the errors of those listed in Table 2, but with larger
standard deviations.
The {\it GaussFit} results were
independently verified using the ORBIT program (Forveille et al. 1999)
which yielded orbital parameters to within the errors of the {\it GaussFit}
solution. Assuming a mass of $M$ $=$ 0.85 $M_\odot$ for
$\epsilon$~Eri (Drake \& Smith 1993) results in a
$M$\,sin\,$i$ $=$ 0.86 $M_{Jupiter}$. 
Figure~\ref{orbit} shows the robust estimation fit of the orbital solution 
RV measurements including the  linear trend.

\begin{deluxetable}{crr}
\tablecaption{Orbital  Elements of the Planet Around Epsilon Eridani}
\tablewidth{0pt}
\tablehead{
\colhead{Element}    &
 \colhead{Robust Estimation} & \colhead{Least Squares}  }
\startdata
Period  (days) & 2502.1 $\pm$  20.1  & 2503.5 $\pm$  27.1 \nl
T (JD) & 2449194.8  $\pm$ 13.5 &  2449180.62 $\pm$ 17.2 \nl
eccentricity & 0.608 $\pm$ 0.041 & 0.584 $\pm$ 0.043 \nl
$\omega$ (deg) & 48.9 $\pm$ 4.1 & 45.5 $\pm$ 4.3  \nl
K1 (m\,s$^{-1}$) & 19.0   $\pm$1.7 & 19.0  $\pm$ 1.6 \nl
$f(m)$ (solar masses)  & (0.886$\pm$0.248)$\times 10^{-09}$  &
(0.943 $\pm$ 0.266)$\times 10^{-09}$  \nl
Slope (m\,s$^{-1}$\,yr$^{-1}$ )& 0.42  $\pm$ 0.20  &  0.44 $\pm$ 0.19  \nl
\enddata
\end{deluxetable}

The rms scatter of the various data sets about the
orbital curve are listed in Table 1. The average $\sigma$ is about
14 m\,s$^{-1}$, excluding the O$_2$ data. (The telluric method suffers
from larger systematic errors such as winds, pressure, and temperature
changes in the Earth's atmosphere, etc.)
A large part of the scatter may be
related to magnetic activity. Saar \& Donahue (1997) noted that stars as
active as $\epsilon$~Eri should show activity-related 
RV ``jitter'' of 12--17 m\,s$^{-1}$. The scatter of our data is quantitatively
consistent with expectations for star with the same
age and activity level as $\epsilon$~Eri.

\section{The Nature of the RV Variations: Ca II H\& K measurements}

	Compared to the sun,
$\epsilon$~Eri is a modestly active star and the 7 yr
RV period is well within the range
of periods expected for activity cycles.
The Ca II H \& K S-index has proved to be a powerful technique for
discerning activity cycles and rotational modulation in 
magnetically active stars (see Baliunas et al. 1995 and references therein).
If the RV variations of $\epsilon$~Eri were due to stellar activity then one
would expect to see the same period in the Ca II S-index variations which
would exclude the planet hypothesis. 

Observations of the relative flux in the Ca II H and K 
emission cores in $\epsilon$~Eri have been made since 1966 at 
Mount Wilson Observatory with its 100-inch
and 60-inch telescopes. Spectrophotometric records of about 100 lower
main sequence stars show good long-term precision (rms $=$ 1-2\%)
and are described in a series of papers cited in Baliunas
et al. (1995). Although the program of observations began with measurements
made approximately monthly during the seasonal window of accessibility of each
star for observations, by 1980 observations were made more frequently in order
to record modulation of the Ca II flux by axial rotation owing to the
uneven longitudinal distribution of Ca II emitting regions.

The interval of Ca II spectrophotometric measurements that overlap with the
RV data is approximately 1980 through 1999. The Ca II data were analyzed by
one of us (SLB) for sinusoidal periodicities according to the prescription
for unevenly sampled data given in Horne and Baliunas
(1986), and the results given in Table 3.
The periodogram analysis was made on data averaged over 30-day intervals.
The lower panel of Figure 1 shows the Lomb-Scargle periodogram  of the 30-day
S-index averages.

\begin{deluxetable}{lccc}
\tablewidth{30pc}
\tablecaption{Ca II Analysis}
\tablehead{
\colhead{Value }  &  \colhead{Peak 1} & \colhead{Peak 2} & \colhead{Peak 3}}
\startdata
Frequency (c/d)    & 0.049 $\pm$ 0.004&  0.334 $\pm$ 0.004 &  0.265 $\pm$ 0.005 \nl
Period (yr)          & 20.38            & 2.99         & 3.77      \nl
Power                & 18.5             & 15.6         & 12.8     \nl
FAP (\%)              & 3.9 $\times 10^{-5}$ & 0.0006   & 0.01    \nl
Amplitude            & 0.02             & 0.02         & 0.02    \nl
\enddata
\end{deluxetable}

Two significant periodicities, one of ~20 years (FAP $=$ 4$\times 10^{-7}$)
and one close to 3 years (FAP $=$ 6$\times 10^{-6}$) 
are present suggesting multi-mode
variations in surface magnetism, a not uncommon class of Ca II variability
in younger stars like $\epsilon$~Eri. Both periods show strong and
statistically significant normalized power 
in the periodogram. A third, weaker period of $\sim$ 3.8 years
(FAP $\sim$ 0.0001)  is also possibly present.

	The peak in the Ca II periodogram closest to the RV period  is at 
2476 d (= 6.78 yrs;  $\nu$ = 0.147 $\pm$ 0.008).
However, this is only the sixth highest peak in the periodogram and it has a
very high false alarm probability (FAP = 16\%) indicating the period's 
relative insignificance. Furthermore, a detailed investigation 
yielded no convincing correlation between the RV and Ca II
data sets.
We conclude that there
is no significant periodicity in the Ca II spectrophotometric 
measurements near that of the Doppler RV variation. 

\section{Discussion}

	We have provided convincing evidence for a 6.9 yr 
period in the radial velocity variations of $\epsilon$~Eri based on 
six independent data sets taken at four telescopes using
three different measurement techniques.   Although these RV variations
can be well fit with a Keplerian orbit, there might be some concern
that RV variability may in fact be related to the high magnetic activity
of this star.

A periodogram analysis of
contemporaneous Ca II H\&K S-index measurements revealed 
no significant periods at the one found in the RV analysis. This
seems to exclude stellar activity as the
cause of the observed RV variations.  Since  RV measurements
provide an indirect planet detection and the exact relationship
between Ca II and RV variability is poorly understood, we cannot
absolutely exclude that photospheric motion 
related to magnetic
activity  is indeed  responsible for the observed RV variations. 
However, this motion would have to be unrelated to the Ca II 
variability and {\it known} magnetic timescales
on this star 
which is counter to our current understanding of stellar activity.
Furthermore, it is not clear whether
activity related RV variations can mimic Keplerian motion with a high
orbital eccentricity over such a long time scale. 
Given all the available evidence
the planet hypothesis is 
the simplest, most likely explanation for the RV variability of this star.

	This extra-solar planet is particularly
interesting in light of the dusty ring around $\epsilon$~Eri. This ring 
is non-uniform which may be evidence for other planets
in the system (Greaves et al. 1998; Liou \& Zook 1999), most likely
at large orbital radii ($\sim$ 10--30 AU). 
It is not clear if the planet
that we have detected can dynamically influence the dust ring, although
inner planets may play some role in clearing out the inner parts of the
dust disk (Liou \& Zook 1999). Interestingly, we do detect a linear
acceleration of the central star of about 0.4 m\,s$^{-1}$\,yr$^{-1}$
which is consistent with a 2 $M_{Jupiter}$ planet orbiting at the
inner radius of the ring (30 AU).
Such a planet would produce an astrometric
acceleration of 0.03 mas\,yr$^{-2}$ which can easily be confirmed 
by future space-based interferometry missions.  Epsilon Eri may be a good
candidate star for multiple planets and is deserving of future studies.

Adopting a stellar inclination of 
$i$ = 30{\deg} which is consistent with
$V$ sin $i$ measurements (Saar \& Osten 1997)
and the ring inclination  (Greaves et al. 1998)
and assuming that the orbital axis of the planet, 
stellar rotation axis,  and ring axis are all aligned,
we estimate the true mass of the planet to be  1.7  $M_{Jupiter}$.

	The candidate planet around 
$\epsilon$~Eri would qualify as the first ``Jupiter analog'' except for its 
high orbital eccentricity. Giant planets in eccentric orbits
is a common phenomenon among extra-solar planets
(Cochran et al. 1997; Korzenik et al. 2000; 
Vogt et al. 2000). Explanations include 
interactions with a binary companion (Holman et al. 1997),
interaction with the disk (Artymowicz 1992), 
or mergers and scatterings between two or more giant planets 
(e.g. Lin \& Ida 1997).

It seems unlikely that the eccentricity is due to interactions
with a binary companion as there is scant evidence  
for $\epsilon$~Eri being a binary star. The claim of an astrometric
perturbation to $\epsilon$~Eri with a 25 yr period by van de Kamp (1974)
has been refuted by Heintz (1993).
Wielen et al. (1999) compared
ground-based proper motion measurements of $\epsilon$~Eri to those
of {\it Hipparcos} and found that this star was a binary ``on the verge
of detectability and significance''. The ``best-fit'' orbit does have a slight
linear trend which may be due to a long-period companion, but this is not
compelling given that the
slope amounts to a velocity change of
less than 10 m\,s$^{-1}$ over the course of our
measurements or comparable to the RV rms scatter.
Data covering a longer time base is needed to confirm this.
At the present time there is no strong evidence to support the hypothesis
that the high eccentricity is caused
by interactions with a stellar companion. Given the diversity of planets
showing highly eccentric orbits (now with long orbital periods)
there is  probably a number of mechanisms at work producing the
observed eccentricities of extra-solar planets.

	With a period of nearly 7 years and a $K$-velocity amplitude of only
19 {\ms}, the planet of $\epsilon$~Eri has the longest orbital period 
and one of  the lowest velocity amplitudes (for the central star) yet found.
The rms scatter of the RV measurements about the orbital solution 
(due to measurement error and stellar activity) is comparable
to the orbital $K$-amplitude. In spite of this
the planet was detected with a high degree of confidence.
This demonstrates that RV signals 
with amplitudes comparable to the noise level can be detected  reliably 
given sufficient data.   Ancillary measurements such
as Ca II H\&K emission measures are also extremely
important for confirming planet discoveries, particularly long period ones.

The stellar distance of 3.2 pc yields  an 
angular separation between star and planet of about 1 arcsecond. There is
hope that we may 
one day be able to detect light from the planet using ground-based imaging with 
adaptive optics or space-based imaging.
We searched the {\it Hubble Space Telescope} archives and found several
images taken at 2$\mu$m with the coronograph of NICMOS, but no nearby 
companions were seen.  However, 
the albedo of the planet at this wavelength is expected to be
zero (Sudarsky et al. 2000).  Furthermore, 
with an estimated
effective temperature $\sim$ 100 K the radiated flux from the
planet would also be undetected by NICMOS in this bandpass. 
To detect the thermal radiation from the planet requires observations
at wavelengths near 20$\mu$m where the
companion would have a magnitude $\approx$ 20. Searches for reflected
light from the planet must be done in spectral regions where the 
planet albedo is expected to be high.
For example, the theoretical 
albedo at 5500 {\AA} for a gaseous giant planet with 
an effective temperature near 100 K is about 0.7 (Sudarsky et al. 2000).
This results in a  ``reflected'' magnitude for  the planet 
about 21--22  magnitudes fainter than the primary star.

	The astrometric perturbation of the central star 
is estimated to be $\approx$ 2 mas. This is at the limit
of ground based measurements for this star (see Wielen et al. 1999)
but should be 
measurable using  space-based astrometric measurements with the
{\it Hubble Space Telescope} (McArthur et al. 1999)  or 
with future space-based or ground-based (Keck, VLTI)
astrometric measurements.
These combined with the RV measurements should yield  orbital inclination
and thus the true mass of the companion. We note that either
direct imaging or astrometric measurements would provide the best confirmation 
of this  planet.

	The discovery of the planet around
$\epsilon$~Eri has now begun to push the parameter space of extra-solar
planet discoveries to the long period, solar-system giant planet analogs. 
As RV searches
lengthen their time base more of these systems will undoubtedly be discovered.

	A Hatzes and W. Cochran  acknowledges the support of NASA grants
NAG5-4384  and NAG5-9227 and by NSF grant AST-9808980.
B. McArthur  receives support through NASA grant NAS5-1603.
M. Endl acknowledges support by the Austrian Fond zur F\"orderung der
wissenschaftlichen Forschung Nr.~S7302.
S. Baliunas is grateful to the dedication of M. A. Bradford, R. A. Donahue,
K. Palmer and W. H. Soon to the HK Project at Mount Wilson Observatory, which
is operated by the Mount Wilson Institute under
an agreement with the Carnegie Institution of Washington. Support of this
research has been made possible through the generosity of the Richard C.
Lounsbery Foundation.  R.P. Butler and G. Marcy  thank D. Fischer, 
E. Williams, S. Vogt, and C. McCarthy.
Finally, we thank
K. Dennerl and S. D\"obereiner for the help with the ESO observations.

\end{document}